\documentclass{sig-alternate-per}
\usepackage{amsmath}
\usepackage{amsfonts}

\def \e       {{\tt e}}
\def \P       {{\Psi}}
\def \Phead   {{\Psi_{\rm head}}}
\def \Ptail   {{\Psi_{\rm tail}}}
\def \O       {{\Theta}}
\def \Ohead   {{\Theta_{\rm head}}}
\def \Otail   {{\Theta_{\rm tail}}}
\def \Fhead   {{F_{\rm head}}}
\def \Ftail   {{F_{\rm tail}}}
\def \fhead   {{f_{\rm head}}}
\def \ftail   {{f_{\rm tail}}}
\def \ahead   {{a_{\rm head}}}
\def \bhead   {{b_{\rm head}}}
\def \chead   {{c_{\rm head}}}
\def \atail   {{a_{\rm tail}}}
\def \btail   {{b_{\rm tail}}}
\def \ctail   {{c_{\rm tail}}}

\begin{document}


\title{A simple model for citation curve}
\numberofauthors{3}
 \author{
         \alignauthor
         Y.C. Tay\\
         \affaddr{Dept. of Computer Science}\\
         \affaddr{National Univ. of Singapore}\\
         \affaddr{Singapore}\\
         \alignauthor
         Mostafa Rezazad\\
         \affaddr{School of Computer Science}\\
         \affaddr{Institute for Research in Fundamental Sciences (IPM)}\\
         \affaddr{Tehran, Iran}\\
         \alignauthor
         Hamid Sarbazi-Azad\\
         \affaddr{Sharif Univ. of Technology and}\\
         \affaddr{Institute for Research in Fundamental Sciences (IPM)}\\
         \affaddr{Tehran, Iran}\\
         }


\maketitle


\begin{abstract}
There is considerable interest in the citation count for an author's 
publications.
This has led to many proposals for citation indices
for characterizing citation distributions.
However, there is so far no tractable model to facilitate the analysis
of these distributions and the design of these indices.
This paper presents a simple equation for such design and analysis. The equation has three parameters that are calibrated 
by three geometrical characteristics of a citation distribution.
Its simple form makes it tractable.
To demonstrate, the equation is used to derive closed-form expressions
for various citation indices, analyze the effect of time
and identify individual contribution to the Hirsch index for a group.

\end{abstract}


\section{Introduction}

Since the launch in 2004 of the web search engine 
{\sl Google Scholar}\footnote{\tt https://scholar.google.com/},
it has become easy to look for the papers and publication record
of a researcher.
The information provided currently includes the citations for each publication,
and a $h$-index for the citation count.

The $h$-index was proposed by Hirsch in 2005~\cite{Hirsch2005}.
It gained much attention and triggered numerous proposals
for alternative citation indices~\cite{BihariSurvey2018},
but there is controversy over characterizing an author's research record 
by such indices~\cite{CiteStats2009}.

In this paper, we do not advocate one index or another, 
and propose none ourselves.
Instead, we offer a simple equation for approximating the distribution
of citation count.
We claim that this equation can facilitate the analysis of these distributions,
and the design of citation indices.

To demonstrate our claim, we use the equation to derive closed-form
expressions for various indices in Sec.~\ref{sec:related}.
In Sec.~\ref{sec:2appl}, we further apply the equation to examine
the effect of time, and how individual $h$-indices contribute
to the $h$-index of a group of researchers.

\newpage
\section{The proposed model}
\label{sec:equation}

Let $\Psi(n)$ denote the number of citations for an author's $n$-th 
publication, where the publications are sorted based on their citation numbers so that
$\Psi(n)\ge\Psi(n^\prime)$ for $n<n^\prime$.
Let $M=\Psi(1)$, i.e. the maximum number of citations for the author's most cited
publication.
Suppose the author has $N$ cited publications, so $\Psi(N)>0$ but $\Psi(N+1)$
is either 0 or undefined.
We seek a closed-form expression to define a function $f$ 
that approximates $\Psi$.

A frequently-used expression~\cite{Egghe2006} is the power law
\begin{equation}
f(x)=\frac{C}{x^\lambda} ,
\label{eq:power}
\end{equation}
where $C$ and $\lambda$ are parameters that vary among authors,
$C>0$ and $\lambda>0$.
This $f:\mathbb{R}^+\rightarrow\mathbb{R}^+$ has 3 issues:

First, the vertical asymptote at $x=0$ can make the approximation
bad for authors who do not have hugely different citation counts
for top-ranked papers.
Second, the horizontal asymptote as $x\rightarrow\infty$ can give a poor
approximation for authors with a small number of papers $N$.
Third, two parameters do not suffice: 
We can think of $M$ and $N$ as anchoring $f$, 
but they leave much ambiguity for the curvature in between.

(For the power law, if we require $f(1)=M$, then $C=M$ in Eqn.~(\ref{eq:power}),
and curvature is determined by $\lambda$.)

We therefore need at least 3 parameters to specify $f$.
We use the point where $f$ cuts the diagonal line to fix the curvature for $f$, 
i.e. if the real value $h$ is defined by $f(h)=h$,
then $f$ is determined by $M$, $N$ and $h$.

\begin{figure*}
\centerline{
\includegraphics[width=18.3cm]{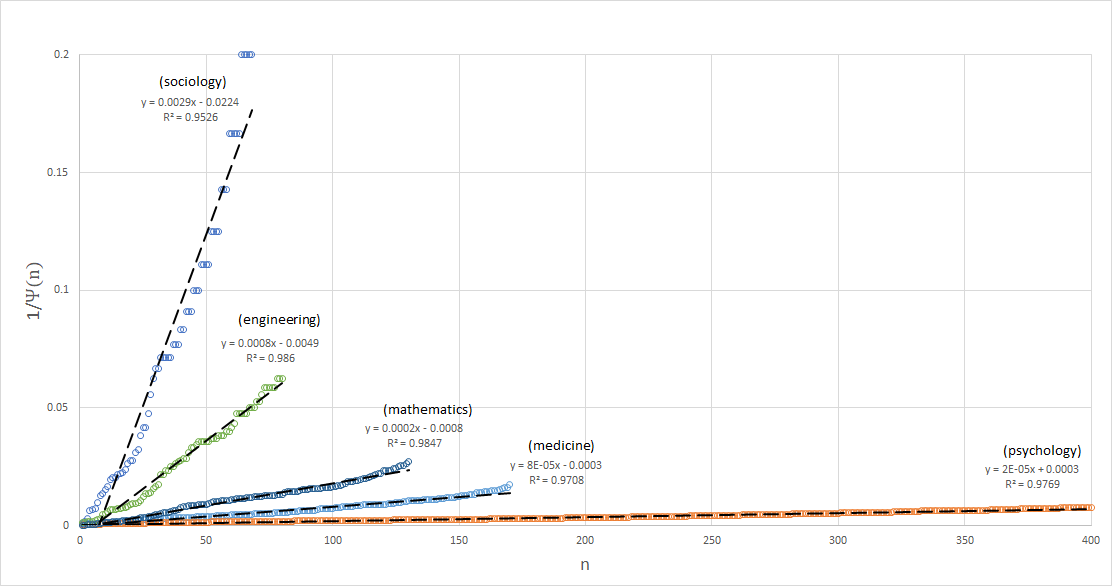}
}
\caption{
Regression lines show that 
$1/\Psi(n)$ is approximately linear for small $n$. 
(The research area for each author is indicated.
Note the nonzero intercepts.)}
\label{fig:reciprocal0}
\end{figure*}

What should we choose for $f$?
For the power law, $1/f(x)$ is proportional to $x^\lambda$.
Fig.~\ref{fig:reciprocal0} plots $1/\Psi(n)$ for 5 researchers
in engineering ($M=718, N=171, h=50$), 
mathematics ($M=8763, N=410, h=77$),
medicine ($M=2019, N=345, h=114$),
psychology ($M=3740, N=982, h=243$) 
and sociology ($M=1057, N=116, h=24$).
The citation data is from a dataset of 226 authors 
that we sampled from {\sl Publish or Perish}\footnote{\tt 
https://harzing.com/resources/publish-or-perish};
the histograms for $M$, $N$ and $h$ are in the Appendix.
We use this dataset throughout this paper.

In each case, the regression line shows that
$1/\Psi(n)$ is approximately linear in $n$ for small $n$,
where most of an author's citations are.
This suggests $f$ should have the form 
$1/f(x)=\gamma_1 x + \gamma_0$ for some constants $\gamma_0$ and $\gamma_1$.
Since $f$ needs to have 3 parameters, we define it as
\begin{equation}
f(x)=\frac{b}{x+c}-a.
\label{eq:f}
\end{equation}
where $a$, $b$ and $c$ are positive real values.

\begin{figure}
\centerline{
\includegraphics[width=8.3cm]{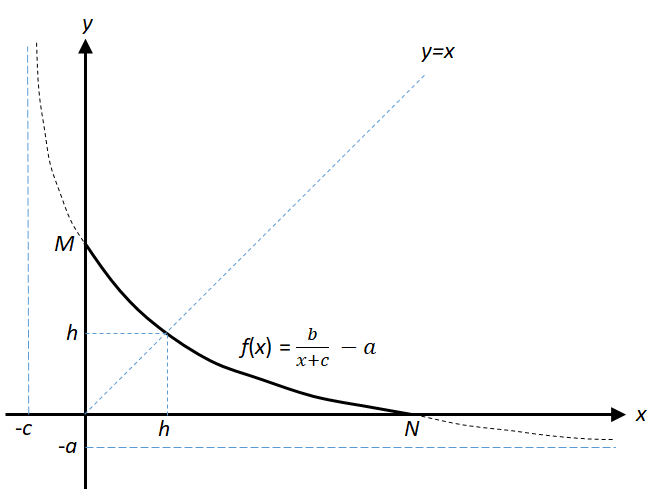}
}
\caption{
The shape for $f(x)$ is determined by $M$, $N$ and $h$,
which calibrate $a$ for the horizontal asymptote, 
$c$ for the vertical asymptote, and $b$ for the curvature.
}
\label{fig:reciprocal1}
\end{figure}

Fig.~\ref{fig:reciprocal1} illustrates this $f$.
The function has a horizontal asymptote at $y=-a$, vertical asymptote at $x=-c$,
and intersects $y=x$ at $x=h$.
These 3 parameters ($a,b,c$) control the location and curvature of $f$.
Their values are determined by
\begin{equation}
f(0)=M,\quad f(N)=0\quad {\rm and }\quad f(h)=h.
\label{eq:calibrate}
\end{equation}
It follows that $f(1)\approx f(0)=M=\Psi(1)$,
and $f(N)=0\approx\Psi(N)$.
Moreover, the $h$-index~\cite{Hirsch2005}
is defined by solving $\Psi(n)=n$, so $f(h)=h\approx\Psi(h)$.
We are only interested in $f(x)$ for $0\le x\le N$.

Solving Eqn.~(\ref{eq:f}) and Eqn.~(\ref{eq:calibrate}) gives
\begin{align}
\begin{split}
a &= \frac{Mh^2}{MN-(M+N)h} \label{eq:abc}\\
b &= MN(M-h)(N-h) \big(\frac{h}{MN-(M+N)h}\big)^2 \\
c &= \frac{Nh^2}{MN-(M+N)h} \\
\end{split}
\end{align}

\begin{figure*}
\centerline{
\includegraphics[width=8.3cm]{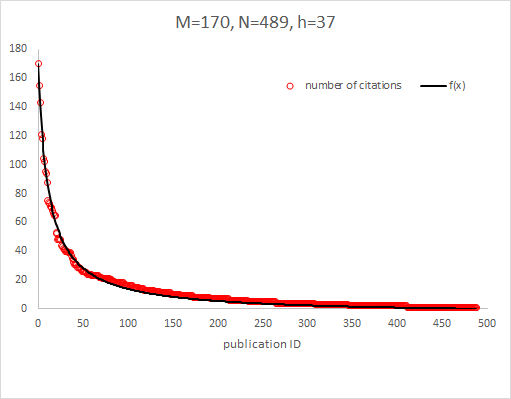}
\quad
\includegraphics[width=8.3cm]{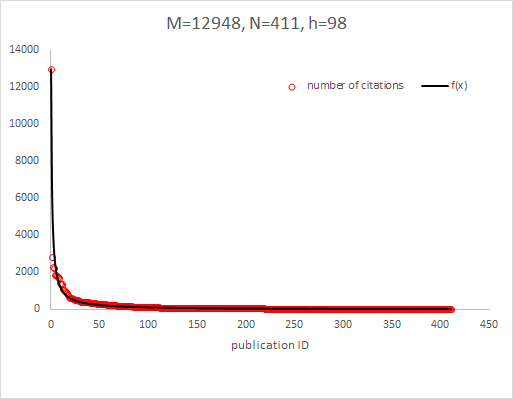}
}
\centerline{\hglue 3.8cm (i) small $M$ 
          \hglue 7.0cm (ii) large $M$ \hfill}

\vglue 20pt
\centerline{
\includegraphics[width=8.3cm]{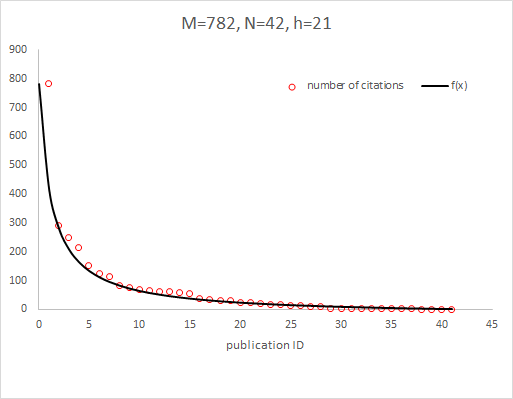}
\quad
\includegraphics[width=8.3cm]{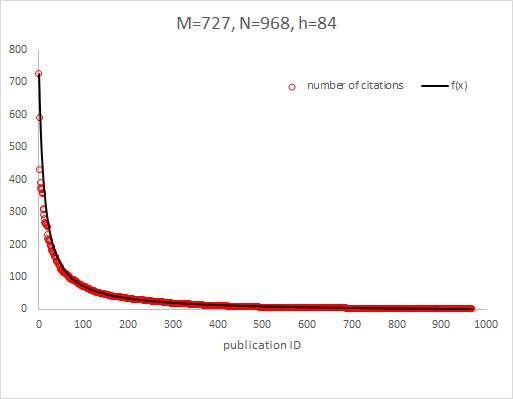}
}
\centerline{\hglue 3.8cm (iii) small $N$ 
          \hglue 7.0cm (iv) large $N$ \hfill}

\vglue 20pt
\centerline{
\includegraphics[width=8.3cm]{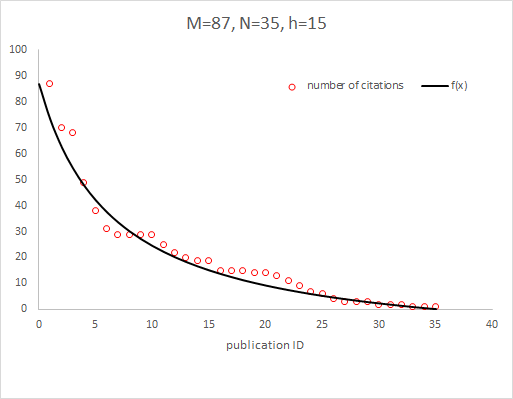}
\quad
\includegraphics[width=8.3cm]{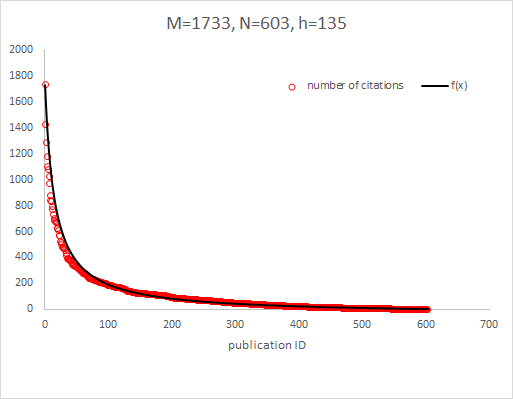}
}
\centerline{\hglue 3.8cm (v) small $h$ 
          \hglue 7.0cm (vi) large $h$ \hfill}

\caption{
Comparing number of citations $\Psi(n)$ and approximation $f(x)$ 
for 6 different authors whose research areas are in computer science, chemistry
and medicine.}
\label{fig:f}
\end{figure*}

Fig.~\ref{fig:f} shows how well $f(x)$ fits $\Psi(n)$
for 6 researchers from our dataset.

\subsection{Approximations for the head and tail}

\begin{figure}
\centerline{
\includegraphics[width=8.3cm]{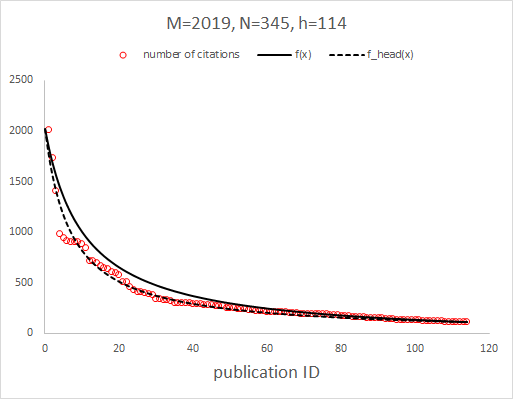}
}
\caption{
$\fhead$ can provide a better fit for $\Psi(n)$, $n=1,\ldots,h$.  (The author is a computational biologist.)
}
\label{fig:Korberfhead}
\end{figure}

Let $\Phead$ and $\Ptail$ denote the $\P$ for $n\le h$ and $n>h$
respectively.
One can simplify the expressions in Eqn.~(\ref{eq:abc})
by focusing on $\Phead$ and neglecting the fit for $\Ptail$.
We do this by taking the limit $N\rightarrow\infty$ in Eqn.~(\ref{eq:abc})
and thus derive another approximation $\fhead$ from $f$:
\begin{align}
\begin{split}
\ahead &= 0 \label{eq:abchead} \\
\bhead &= \frac{Mh^2}{M-h} \\
\chead &= \frac{h^2}{M-h}  \\
{\rm and}\quad \fhead(x) &= \frac{\bhead}{x+\chead}. 
\end{split}
\end{align}
In effect, $\fhead$ is obtained from $\P$ by solving
\[
\fhead(0)=M,\quad \fhead(h)=h\quad{\rm and}\quad 
\lim_{x\to\infty}\fhead(x)=0
\]
using $M$ and $h$ to calibrate the two parameters $\bhead$ and $\chead$
for $\fhead$.
Fig.~\ref{fig:Korberfhead} shows how $\fhead$ can give a better fit 
if $f$ over-estimates $\Psi$.

We can similarly extract an approximation $\ftail$ for $\Ptail$
by taking $M\rightarrow\infty$.
This gives
\begin{align}
\begin{split}
\atail &= \frac{h^2}{N-h} \label{eq:abctail} \\
\btail &= \frac{Nh^2}{N-h} \\
\ctail &= 0 \\
{\rm and}\quad \ftail(x) &= \frac{\btail}{x}-\atail. 
\end{split}
\end{align}
Equivalently, we use $N$ and $h$ to calibrate parameters $\atail$ and $\btail$
in $\ftail$ by solving
\[
\ftail(N)=0,\quad
\ftail(h)=h
\quad{\rm and}\quad 
\lim_{x\to 0}\ftail(x)=\infty .
\]

\subsection{Areas under the curve}

Let $\O$ denote the {\it total} number of citations 
for an author's publications.
Using $f$, we can approximate $\O$ by
\begin{align}
F &= \int_0^N f(x) dx = \int_0^N\big(\frac{b}{x+c} - a\big)dx \nonumber\\
  &= b\ln{\big(1+\frac{N}{c}\big)}-aN \nonumber \\
  &=\frac{MN(M-h)(N-h)h^2}{(MN-(M+N)h)^2}
    \ln{\big(1+\frac{MN-(M+N)h}{h^2}\big)} \nonumber \\
  &\quad -\frac{MNh^2}{MN-(M+N)h}\label{eq:F7} \\
  &\approx h^2\ln{\big(\frac{MN}{\e h^2}\big)}
   \quad{\rm for}\ M>>h\ {\rm and}\ N>>h.\label{eq:F8}
\end{align}
Here, we use $\e$ to denote Euler's number ($\e\approx2.71828$),
to avoid confusion with $e$ in the $e$-index below.

\begin{figure*}
\centerline{
\includegraphics[width=8.3cm]{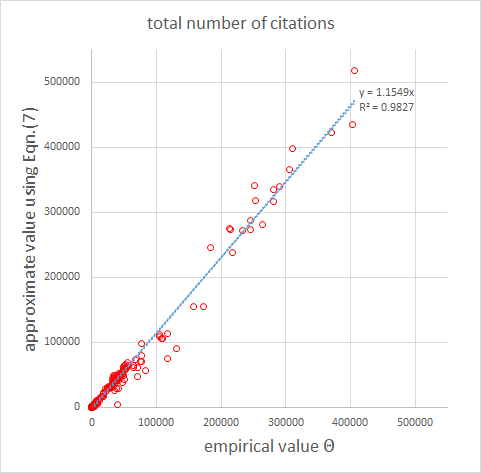}
\qquad
\includegraphics[width=8.3cm]{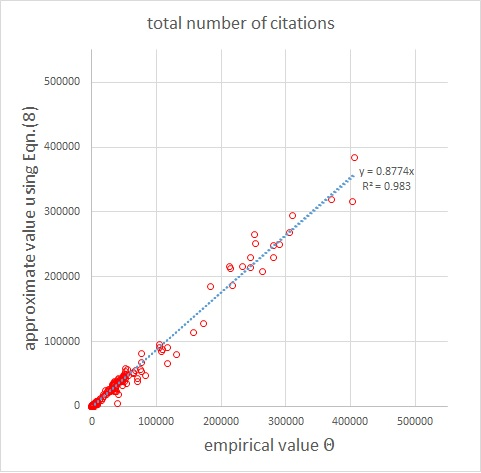}
}
\centerline{
\hglue 1cm (a) Eqn.~(\ref{eq:F7}) over-estimates $F$ (on average)
\hglue 3cm (b) Eqn.~(\ref{eq:F8}) under-estimates $F$ (on average)
\hfill}
\caption{
Total number of citations:
comparing approximate value $F$ to empirical value $\O$.
(In this and the following plots, the regression lines are
constrained to pass through 0.)
}
\label{fig:F}
\end{figure*}

The above approximations for $F$ are plotted against $\O$ in Fig.~\ref{fig:F}
for our dataset of 226 authors;
if $F$ is an accurate estimate for $\O$, 
then the sample points would scatter around the diagonal line $F=\O$.
Fig.~\ref{fig:F} shows that Eqn.~(\ref{eq:F7}) over-estimates $\O$ (on average),
since the regression line has a gradient of 1.15,
whereas Eqn.~(\ref{eq:F8}) under-estimates $\O$ (on average),
since the regression line has a gradient of 0.88.

In Fig.~\ref{fig:F}(a), there is an obvious outlier at \\
$(\O,F)=(39868, 4943)$; 
it is from a computer scientist with an extremely skewed $\Psi$ that drops 
from $\Psi(1)=28142$ to $\Psi(10)=102$.
This author appears as outlier in most of the following plots as well.

One could improve on the approximation by shifting $f$ so
$f(1)=M$ (instead of $f(0)=M$), but that would further complicate 
the expressions for $a,b$ and $c$ in Eqn.~(\ref{eq:abc}).

Let $\Ohead$  denote the total number of citations for 
an author's first $h$ publications.
We can use $\fhead$ to approximate $\Ohead$ by
\begin{align}
\Fhead &= \int_0^h\frac{\bhead}{x+\chead} dx
        = \bhead\ln{\big(1+\frac{h}{\chead}\big)}\nonumber\\
       &=\frac{Mh^2}{M-h}\ln{\big(1+\frac{M-h}{h}\big)}\nonumber\\
       &=\frac{Mh^2}{M-h}\ln{\big(\frac{M}{h}\big)}\nonumber\\
       &=h\left(\frac{\ln{\frac{1}{h}}-\ln{\frac{1}{M}}}
                     {\frac{1}{h}-\frac{1}{M}}\right) \label{eq:Fhead}
\end{align}
Similarly, if $\Otail$ denotes the total number of citations
for the $N-h$ publications in the tail,
then we can use $\ftail$ to approximate $\Otail$ by
\begin{align}
\Ftail &=\int_h^N\big(\frac{\btail}{x}-\atail\big) dx
        =\btail\left(\ln{\frac{N}{h}}\right)-\atail(N-h) \nonumber \\
       &=\frac{Nh^2}{N-h}\left(\ln{\frac{N}{h}}\right)-h^2 .\label{eq:Ftail}
\end{align}

\section{Closed-form expressions for indices}
\label{sec:related}

In this section, we relate $f$, $\fhead$ and $\ftail$ to previous work.  In particular, we use our equations to derive
closed-form expressions for various indices.

\subsection{Total number of citations}
\label{sec:theta}

When introducing the $h$-index, Hirsch~\cite{Hirsch2007} postulated that
the total number of citations 
\begin{equation}
\O = \alpha h^2
\label{eq:ah2}
\end{equation}
for some $\alpha $ that varies among authors and, empirically, $3<\alpha <5$.
Eqn.~(\ref{eq:ah2}) has the equivalent form
\[
h=\frac{1}{\sqrt{\alpha}}\O^{0.5}.
\]
A regression analysis by van Raan~\cite{vanRaan2006} 
using data for chemistry research in Dutch universities also shows
\[ 
h\ =\ 0.42 \O^{0.45},
\]
where the exponent 0.45 is close to 0.5.
(See also the Yong's ``rule of thumb''~\cite{Yong2014}.)
Using our approximation $F$ for $\O$,
Eqn.~(\ref{eq:F8}) shows that, in fact,
\[
\alpha\approx\ln{\big(\frac{MN}{\e h^2}\big)}\ ;
\]
note that $\alpha$ itself depends on $h$.


\subsection{$A$-index}
\label{sec:Aindex}

\begin{figure}
\centerline{
\includegraphics[width=8.3cm]{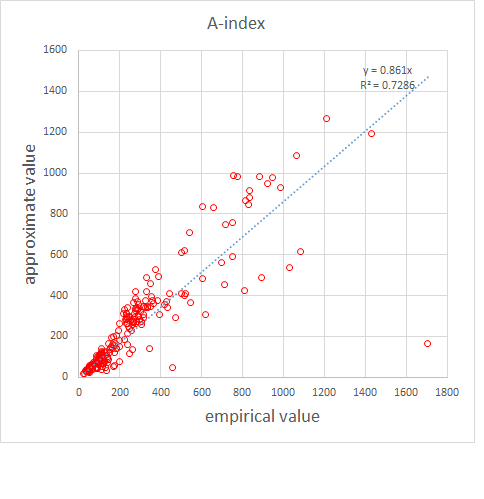}
}
\caption{
Comparing approximate value of $A$-index in Eqn.~(\ref{eq:AF})
to empirical value.
}
\label{fig:Afig}
\end{figure}

To take into account $\Psi(n)-h$ for $n=1,\ldots,h$,
Jin et al.~\cite{Jin2007} defined an $A$-index 
\begin{equation}
A=\Ohead/h ,
\label{eq:AO}
\end{equation}
i.e. the average citation count for the first $h$ papers.
Using Eqn.~(\ref{eq:Fhead}), we can approximate this $A$-index as
\begin{equation}
A\approx\frac{\Fhead}{h}=\frac{Mh}{M-h} \ln{\Big(\frac{M}{h}\Big)}.
\label{eq:AF}
\end{equation}
Fig.~\ref{fig:Afig} compares the empirical value of the $A$-index
for the 226 authors to the approximate value computed with Eqn.~(\ref{eq:AF}).

The regression line has a gradient of 0.86 and correlation coefficient 
$R^2\approx0.72$.
This weak accuracy is expected, since the $A$-index is defined with $\Ohead$,
where there are huge differences among authors in the shape of $\Psi$
for their highly-cited papers.
The outlier at $(1705,164)$ is from the computer scientist previously mentioned
for Fig.~\ref{fig:F}.
Another outlier at $(455,50)$ is from a chemist with just $N=15$ publications,
but has $M=2259$.

Jin et al. proved that, using the power law model $f(x)=M/x^\lambda$,
$A/h$ is a constant determined only by the curvature parameter $\lambda$.
In contrast, Eqn.~(\ref{eq:AF}) shows that, using our model, 
\[
\frac{A}{h}\approx \frac{M}{M-h} \ln{\Big(\frac{M}{h}\Big)},
\]
so $A/h$ also depends on the maximum citation $M$.

\subsection{$R$-index}
\label{sec:Rindex}

\begin{figure}
\centerline{
\includegraphics[width=8.3cm]{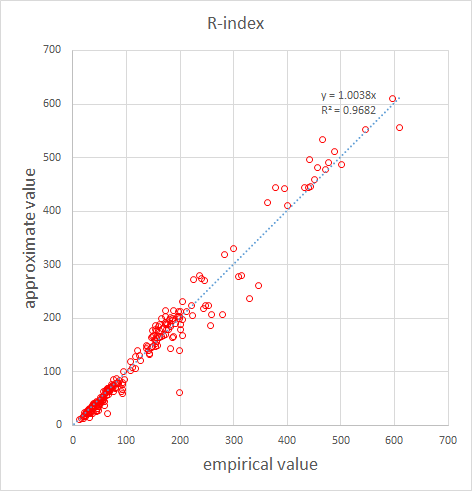}
}
\caption{
Comparing approximate value of $R$-index in Eqn.~(\ref{eq:Rindex})
to empirical value.
}
\label{fig:Rfig}
\end{figure}

As an alternative to the $A$-index,
Jin et al.~\cite{Jin2007} also defined an $R$-index 
\[
R=\sqrt{\Ohead} .
\]
By Eqn.~(\ref{eq:Fhead}), we can approximate this as
\begin{equation}
R\approx\sqrt{\Fhead}=h \sqrt{\frac{M}{M-h} \ln{\Big(\frac{M}{h}\Big)}}\ .
\label{eq:Rindex}
\end{equation}
Fig.~(\ref{fig:Rfig}) shows that, although there are outliers,
Eqn.~(\ref{eq:Rindex}) provides a closed-form expression that, in general,
gives an excellent approximation for the $R$-index
(the regression line has gradient 1.00 and $R^2\approx0.97$).

Again, under the power law, $R/h$ is expressible in terms of $\lambda$,
whereas Eqn.~(\ref{eq:Rindex}) shows that the ratio depends on $M$ as well.


\subsection{$g$-index}
\label{sec:gindex}

An author's $h$-index remains the same no matter how high $\Psi(n)$ is
for $n=1,2,\ldots,h$.
To overcome this issue, Egghe defined a $g$-index~\cite{Egghe2006},
which we approximate as
\begin{equation}
g \approx \frac{\int_0^g f(x) dx}{g} .
\end{equation}
Egghe has shown that $g>h$.
We therefore use $\fhead$ and $\ftail$ to further approximate $g$ by
\begin{align}
g^2 &\approx \int_0^h\fhead(x)dx + \int_h^g\ftail(x)dx \nonumber\\
&=\frac{Mh^2}{M-h}\ln\big(\frac{M}{h}\big) 
  + \btail\ln{\big(\frac{g}{h}\big)} - \atail(g-h) \nonumber\\
&\phantom{XXXXXXXXXX}{\rm by\ Eqn.~(\ref{eq:abctail})
                        \ and\ Eqn.~(\ref{eq:Fhead})} \nonumber\\
&\approx \frac{Mh^2}{M-h}\ln\big(\frac{M}{h}\big) 
       + \frac{Nh^2}{N-h}\ln\big(\frac{g}{h}\big) 
         \quad{\rm for\ }\atail\approx0                \nonumber\\
&\approx h^2\ln\big(\frac{M}{h}\big) 
       + h^2\ln\big(\frac{g}{h}\big) 
         \quad{\rm for\ }M>>h\ {\rm and\ }N>> h. 
         \label{eq:g2}
\end{align}
Consider some $\beta>\frac{g}{h}$.  Then
\begin{align}
\ln\big(\frac{g}{h}\big)        
&= \ln\beta+\ln\big(1-\big(1-\frac{g}{\beta h}\big)\big)\nonumber\\
&\approx\ln\beta - \big(1-\frac{g}{\beta h}\big) \nonumber\\
&\phantom{XXXXXXX}{\rm since\ }\ln(1-x)\approx-x\ {\rm for\ }0<x<1.\nonumber
\end{align}
The citation data we have seen all show $\frac{g}{h}<4$,
so we choose $\beta=4$.
(A larger $\beta$ will increase $1-\frac{g}{\beta h}$
and worsen the approximation.)
Therefore,
\[
\ln\big(\frac{g}{h}\big)\approx \ln\big(\frac{4}{\e}\big) +\frac{g}{4h}\ .
\]
Substituting this into Eqn.~(\ref{eq:g2}\big), we get
\[
\frac{g^2}{h^2}\approx\ln\big(\frac{M}{h}\big)+\ln\big(\frac{4}{\e}\big)
+\frac{g}{4h}\ ,
\]
so
\[
\Big(\frac{g}{h}-\frac{1}{8}\Big)^2-\frac{1}{64}
\approx\ln\big(\frac{4M}{\e h}\big)\ .
\]
Since $\frac{g}{h}>1$, we can further simplify this as
\begin{align}
\Big(\frac{g}{h}\Big)^2 &\approx \ln\big(\frac{4M}{\e h}\big) \nonumber\\
{\rm i.e.}\quad g &\approx h\sqrt{\ln\big(\frac{4M}{\e h}\big)} \label{eq:g3}
\end{align}

Egghe has proven that, for the power law model $f(x)=M/x^\lambda$,
$g/h$ is a constant determined by the curvature parameter $\lambda$ 
only~\cite{Egghe2006}.
In contrast, Eqn.~(\ref{eq:g3}) shows that, for our model,
$g/h$ also depends on $M$.

\begin{figure}
\centerline{
\includegraphics[width=8.3cm]{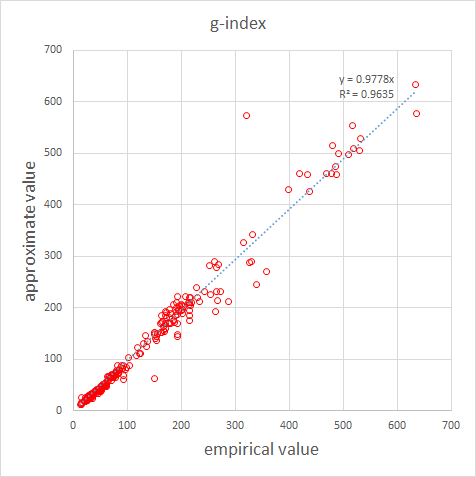}
}
\caption{
Comparing approximate value $g$ in Eqn.~(\ref{eq:g3}) to empirical value $g$.}
\label{fig:gindex}
\end{figure}

Fig.~\ref{fig:gindex} plots the approximation (\ref{eq:g3})
against actual $g$ values for the previously chosen 226 authors.
The regression line shows that, on average, the approximation is accurate.
The under-estimating outlier at $(151, 63)$ 
is from the computer science previously mentioned for Fig.~\ref{fig:F};
the over-estimating outlier at $(320, 573)$ is from an immunologist
with unusually large $M$ and $N$ $(M=12172, N=995, h=281)$, 
for whom $\fhead$ over-estimates $\Phead$.

\subsection{$hg$-index}
\label{sec:hgindex}

\begin{figure}
\centerline{
\includegraphics[width=8.3cm]{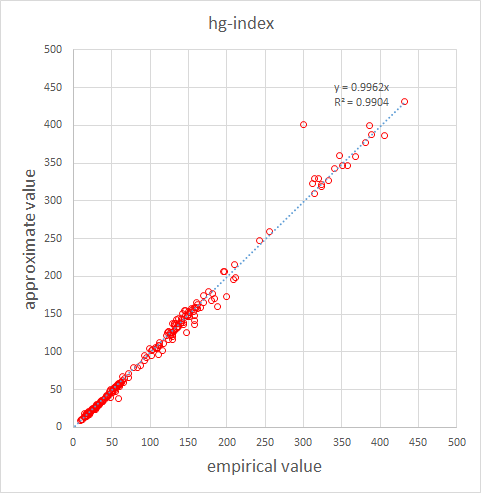}
}
\caption{
Comparing approximate value of $hg$-index in Eqn.~(\ref{eq:hg})
to empirical value.
}
\label{fig:hgfig}
\end{figure}

Alonso et al.~\cite{AlonsoHG2010}
combined the $g$- and $h$-indices to get an hg-index $\sqrt{hg}$.
By Eqn.~(\ref{eq:g3}),
\begin{equation}
\sqrt{hg}\approx h\sqrt[4]{\ln\big(\frac{4M}{\e h}\big)},
\label{eq:hg}
\end{equation}
so the hg-index just reduces the square root in Eqn.~(\ref{eq:g3})
to a 4th root.
Fig.~\ref{fig:hgfig} shows the closed-form accurately approximates
the empirical value of the $hg$-index.

\subsection{$e$-index}
\label{sec:eindex}

\begin{figure}
\centerline{
\includegraphics[width=8.3cm]{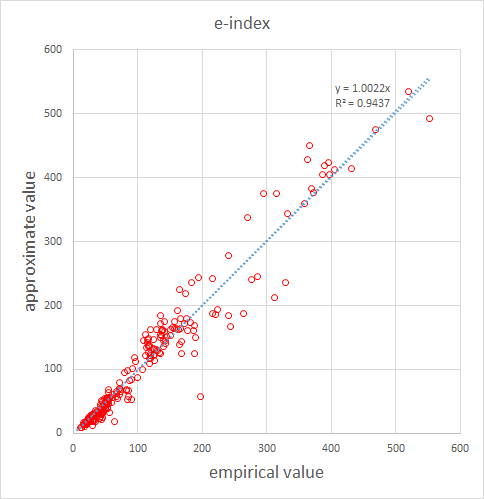}
}
\caption{
Comparing approximate value of $e$ in Eqn.~(\ref{eq:eindex})
to empirical value for the $e$-index.
}
\label{fig:efig}
\end{figure}

Zhang~\cite{Zhang2009} pointed out that the $h$-index holds no information
for the head, and its integer value has coarse granularity.
To address these issues, he proposed the $e$-index, defined by
\[
e^2=\Ohead-h^2.
\]
Using our $\Fhead$ approximation (\ref{eq:Fhead}), we have
\begin{equation}
e^2\approx \frac{Mh^2}{M-h}\ln{\big(\frac{M}{h}\big)}-h^2,
\label{eq:eindex}
\end{equation}
Fig.~\ref{fig:efig} shows the value of $e$ computed from Eqn.~(\ref{eq:eindex}) 
is a good approximation of the empirical value for our dataset: 
The regression line has gradient 1.00 and $R^2\approx 0.94$.

For the power law model $f(x)=M/x^\lambda$, Zhang derived
\[
e^2=M(\frac{1}{2}\ln{M}-1)\quad{\rm if}\ M=h^2,\  
{\rm in\ the\ case\ }\lambda=1.
\]
We can get this from Eqn.~(\ref{eq:eindex}) for the case $M=h^2$ and $h^2>>h$.
In this sense, Zhang's formula for $e$ is a validation
of Eqn.~(\ref{eq:Fhead}) for $f(x)=M/x$ and $M>>h$.

\subsection{$h^\prime$-index}
\label{sec:hprime}

\begin{figure}
\centerline{
\includegraphics[width=8.3cm]{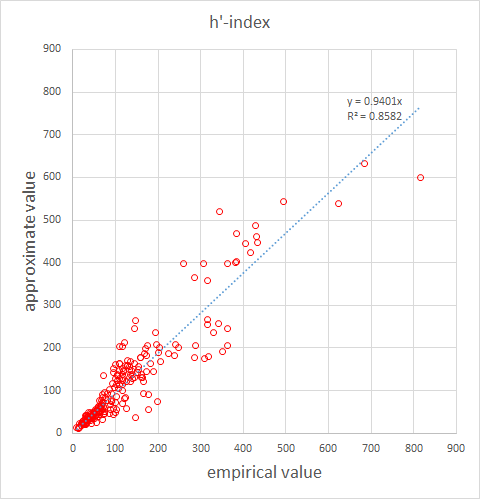}
}
\caption{
Comparing approximate value of $h^\prime$ in Eqn.~(\ref{eq:hprime})
to empirical value for the $h^\prime$-index.
}
\label{fig:hprimefig}
\end{figure}

The $h$-index does not differentiate between authors whose 
$\Ohead$ and $\Otail$ are very different.
To reflect such differences, Zhang defined another index
$h^\prime$~\cite{Zhang2013}, where
\[
h^\prime=\sqrt{\frac{\Ohead-h^2}{\Otail}}\ h.
\]
Using our approximations for $\Ohead$ and $\Otail$, we get
\begin{align}
h^\prime &\approx \sqrt{\frac{\Fhead-h^2}{\Ftail}}\ h\nonumber\\
&= \sqrt{\frac{\frac{M}{M-h}\left(\ln{\frac{M}{h}}\right)-1 }
        {\frac{N}{N-h}\left(\ln{\frac{N}{h}}\right)-1 }}\ h 
\label{eq:hprime}
\end{align}
Fig.~\ref{fig:hprimefig} shows a regression line of gradient 0.94
for approximate $h^\prime$ values calculated with Eqn.~(\ref{eq:hprime})
plotted against the empirical values.
This is a good approximation for expected $h^\prime$ value, 
but the data points are quite dispersed,
giving a correlation coefficient of $R^2\approx0.86$.

For authors with $M>>h$ and $N>>h$, this simplifies to
\begin{align}
h^\prime 
&\approx \sqrt{\frac{\ln{\frac{M}{h}}-1 }{\ln{\frac{N}{h}}-1 }}\ h \nonumber\\
&= \sqrt{\frac{\ln{M}-\ln{\e h}}{\ln{N}-\ln{\e h}}}\ h 
\label{eq:hprimeA}
\end{align}
(Note: It is not uncommon for authors to have $M<\e h$ or $N<\e h$,
for whom the assumption $M>>h$ or $N>>h$ is violated 
and the approximation fails.)

\begin{figure}
\centerline{
\includegraphics[width=8.3cm]{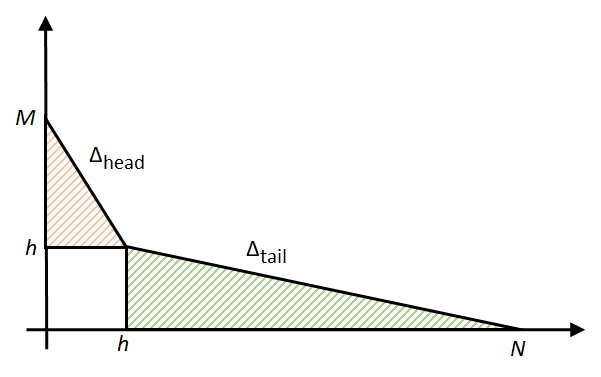}
}
\caption{
Understanding Eqn.~(\ref{eq:hprimeA}) with a simplified geometry.
}
\label{fig:2triangles}
\end{figure}

To better understand the expression in Eqn.~(\ref{eq:hprimeA}), 
consider a simplified geometry using two triangles, as shown in 
Fig.~\ref{fig:2triangles}, with areas
\[
\Delta_{\rm head}=\frac{1}{2}(M-h)h\quad {\rm and}\quad 
\Delta_{\rm tail}=\frac{1}{2}(N-h)h .
\]
For this geometry, we get
\[
h^\prime=\sqrt{\frac{\Delta_{\rm head}}{\Delta_{\rm tail}}}\ h
        =\sqrt{\frac{M-h}{N-h}}\ h.
\]
Eqn.~(\ref{eq:hprimeA}) shows how $h^\prime$ for this simplified geometry
is modified when we take into account the citation curvatures.

\subsection{$h_2$-index}
\label{sec:h2}

In analyzing the citations for an author's publications,
one must first filter out those by another author with a similar name.
(E.g. The Computer Science bibliography website DBLP lists more than 300
authors named ``Wei Wang''.)

\begin{figure}
\centerline{
\includegraphics[width=8.3cm]{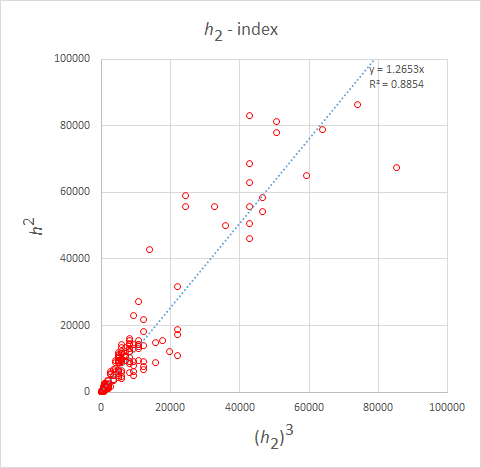}
}
\caption{
$h^2$ over-estimates $h_2^3$.
}
\label{fig:h2fig}
\end{figure}

To reduce the effort needed to disambiguate authorship,
Kosmulski defined a $h_2$-index as the greatest integer such that
the $h_2$ most-cited papers have at least $h_2^2$ citations each.
He observed that
\begin{equation}
h_2\ {\displaystyle\propto}\ \O^{\frac{1}{3}}
\label{eq:h2a}
\end{equation}
The focus of $h_2$ is in the head, so we can approximate $h_2$ by
\begin{align}
h_2^2 &\approx \fhead(h_2) \nonumber\\
      &=\frac{\bhead}{h_2+\chead} \nonumber\\
      &=\frac{Mh^2}{h_2(M-h)+h^2}
                   \quad {\rm by\ Eqn.~(\ref{eq:abchead})}\nonumber\\
      &\approx\frac{Mh^2}{h_2 M+h^2}\quad {\rm for\ }M>>h \nonumber
\end{align} 
Therefore,
\begin{align}
h_2^3M &= h^2(M-h_2^2) \nonumber\\
{\rm i.e.}\quad h_2^3 &=\frac{M-h_2^2}{M} h^2 \nonumber\\
{\rm so}\quad h_2^3   &\approx h^2 \quad {\rm for\ }M>>h_2^2 .
\label{eq:h2index}
\end{align}
This last approximation makes $h^2$ an over-estimate of $h_2^3$, 
as indicated by the regression line (gradient 1.27) in Fig.~\ref{fig:h2fig}.

Eqn.~(\ref{eq:h2index}) thus confirms Kosmulski's observation 
(from a small dataset) that $h_2^3=O(h^2)$.
Using Eqn.~(\ref{eq:F8}), we get
\[
h_2^3\approx \frac{F}{\ln\big(\frac{MN}{\e h^2}\big)} \quad{\rm so\ }
h_2\approx \Big(\frac{1}{\ln\big(\frac{MN}{\e h^2}\big)}\Big)^{\frac{1}{3}}  
\Theta^{\frac{1}{3}}
\]
since $F$ is an approximation for $\Theta$.
We thus see how Kosmulski's approximation (\ref{eq:h2a})
depends on $M$, $N$ and $h$.

\subsection{$dc_i$ and $dc_o$: impact and potential}
\label{sec:impactpotential}

\begin{figure*}
\centerline{
\includegraphics[width=8.3cm]{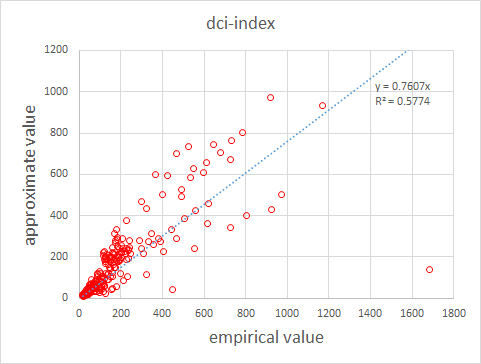}
\qquad
\includegraphics[width=8.3cm]{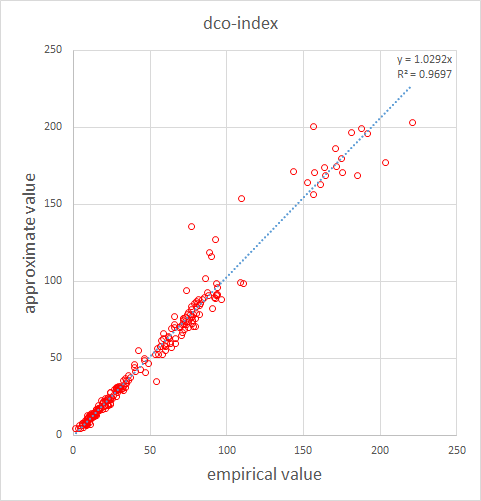}
}
\centerline{
\hglue 2cm
(a) Approximate vales using Eqn.~(\ref{eq:dci1}).
\hglue 3cm
(b) Approximate vales using Eqn.~(\ref{eq:dco1}).
\hfill
}
\caption{
Comparing approximate and empirical values for $dc_i$ and $dc_o$.
}
\label{fig:dcfig}
\end{figure*}

To measure the impact of an author's publications, Silva and Gr\'{a}cio
defined an index~\cite{Silva2021}
\[
dc_i=\frac{1}{h}\sum_{n=1}^h(\Psi(n)-h),
\]
i.e. $dc_i= e^2/h$, using the $e$-index.
From the approximation for $e$ in Sec.~\ref{sec:eindex},
we get
\begin{equation}
dc_i \approx h(\frac{M}{M-h}\ln\frac{M}{h}-1) 
\label{eq:dci1}
\end{equation}
Fig.~\ref{fig:dcfig}(a) shows that the agreement between this approximation
and the empirical $dc_i$ values is weakened by dispersion ($R^2\approx0.58$)
and outliers (gradient$\approx0.76$).
As with the $A$-index in Fig.~\ref{fig:Afig},
it is hard to give an accurate formula for $dc_i$
since authors have large differences in citation patterns 
for their highly-cited papers.

The motivation for $dc_i$ lies in differentiating two authors with similar $h$
value. 
From Eqn.~(\ref{eq:dci1}), we get
\begin{equation}
dc_i \approx h \ln\frac{M}{\e h}\quad{\rm for\ }M>>h, 
\label{eq:dci2}
\end{equation}
and we see that the author with a larger $M$ indeed has a higher $dc_i$.
Silva and Gr\'{a}cio's citation data for 116 Brazilian mathematicians
has a moderate Pearson correlation coefficient of 0.48 between $dc_i$ and $h$,
and we see this correlation in Eqn.~(\ref{eq:dci2})
if $\ln\frac{M}{\e h}$ is considered a multiplicative noise term for fixed $h$.

To measure the potential for increasing an author's $h$ value,
Silva and Gr\'{a}cio defined an index
\[
dc_o=\frac{1}{N-h}\sum_{n>h}(h-\Psi(n)),
\]
i.e. $dc_o=h-\Otail/(N-h)$.
Using $\Ftail$ in Eqn.~(\ref{eq:Ftail}) to approximate $\Otail$, we get
\begin{equation}
dc_o \approx h-\frac{1}{N-h}\big(\frac{Nh^2}{N-h}\ln{(\frac{N}{h})} - h^2\big)
\label{eq:dco1}
\end{equation}
Fig.~\ref{fig:dcfig}(b) shows that
--- in contrast to Fig.~\ref{fig:dcfig}(a) --- 
there is excellent agreement between this approximation
and the empirical value (gradient$\approx1.03$, $R^2\approx0.97$).

Silva and Gr\'{a}cio argue that, for two authors with the same $h$ value,
the one with a larger $dc_o$ value has greater potential for increasing $h$.
We can see this by further approximating Eqn.~(\ref{eq:dco1}):
\begin{align}
dc_o \approx h\big(1-\frac{h}{N}\ln(\frac{N}{\e h})\big),
\label{eq:dco2}
\end{align}
where, for the same $h$, $dc_o$ is higher for the author with a larger $N$.
We can rewrite Eqn.~(\ref{eq:dco2}) as
\[
dc_o\approx h - \frac{h^2}{N}\ln(\frac{N}{\e h}),
\]
so $dc_o$ is linear in $h$, with an additive noise term 
$\frac{h^2}{N}\ln(\frac{N}{\e h})$
induced by $N$ and biased by $h$.
Indeed, Silva and Gr\'{a}cio's data shows a strong linear correlation 
(Pearson coefficient $\approx 0.96$)
between $dc_o$ and $h$.

\section{Considering time and group \\ activities}
\label{sec:2appl}

We now apply our approximations to analyze the effect of time
and the aggregation of citation counts.

\subsection{Modeling the effect of time}
\label{sec:time}

In using $dc_o$ to measure the potential for increasing an author's $h$ value,
one can make a prediction.
Supporting such a prediction requires some model of how citations and
publications increase over time.

When introducing the $h$-index, Hirsch gave a back-of-an-envelope derivation
that shows
\begin{equation}
h=h_0 t ,
\label{eq:h0}
\end{equation}
where $t$ is the time since the author's first publication,
and $h_0$ is a constant determined by publication rate and citation rate.
There is some empirical validation of 
Eqn.~(\ref{eq:h0})~\cite{Abt2012,Burrell2007b,Hirsch2007,Liang2006}.

Burrell also provided numerical support using a stochastic 
model~\cite{Burrell2007a}.
This model assumes that the number of publications for an author is Poisson
distributed over time at a constant rate.
One therefore expects
\begin{equation}
N=N_0 t ,
\label{eq:N0}
\end{equation}
for some constant $N_0$.
By a similar Poisson assumption, the number of citations for a particular 
publication is expected to be linear with respect to time.
An author's publications appear at different times and have different
citation rates (that are gamma distributed in Burrell's model).
The publication with the highest citation 
--- and the corresponding citation rate ---
may therefore change over time.
Even so, we further assume
\begin{equation}
M=M_0 t 
\label{eq:M0}
\end{equation}
for some constant $M_0$.
In the following, we refer to Eqns.(\ref{eq:h0})--(\ref{eq:M0})
as the {\it linear model}.

It follows from this model and Eqn.~(\ref{eq:abc}) that
\begin{align}
a &= \frac{M_0 h_0^2}{M_0 N_0 -(M_0 +N_0 )h_0} t \nonumber\\
b &= M_0 N_0 (M_0 -h_0)(N_0 -h_0) 
    \Big(\frac{h_0}{M_0 N_0 -(M_0 +N_0 )h_0}\Big)^2 t^2 \nonumber\\
c &= \frac{N_0 h_0^2}{M_0 N_0 -(M_0 +N_0 )h_0} t .\nonumber
\end{align}
Similarly, $\chead$ and $\atail$ are linear in $t$,
but $\bhead$ and $\btail$ are quadratic in $t$.

It follows from Eqn.~(\ref{eq:g3}) that
\[
g\approx \left(h_0\sqrt{\ln\Big(\frac{4M_0}{\e h_0}\Big)}\right) t .
\]
Burrell observed this linearity in two numerical examples for his 
stochastic model~\cite{Burrell2009}.
He pointed out that the correlation coefficient for $h_2$ vs $t$ 
is much smaller.
In fact, we see from Eqn.~(\ref{eq:h2index}) that 
\[
h_2\approx h_0^{\frac{2}{3}} t^{\frac{2}{3}} ,
\]
so $h_2$ is not linear in $t$.

Burrell's numerical examples also showed that $\Ohead$ and the $A$-index
are approximately proportional to $t^2$ and $t$ respectively.
Indeed, we see from Eqn.~(\ref{eq:Fhead}) and Eqn.~(\ref{eq:AF}) that
\begin{equation}
\Ohead \approx \Fhead \approx 
       \big(\frac{M_0 h_0^2}{M_0 -h_0}\ln{\big(\frac{M_0}{h_0}\big)\big)}t^2
\label{eq:t2}
\end{equation}
and
\[
A \approx
       \left(\frac{M_0 h_0}{M_0 -h_0}\ln{\big(\frac{M_0}{h_0}\big)}\right)t ,
\]
so the multiplicative factors are constant if $M,N$ and $h$ are linear in $t$.
Similarly, the factors
\begin{align}
\ln{\big(\frac{MN}{\e h^2}\big)}
   \quad &{\rm for\ }\O\ {\rm in\ Eqn.~(\ref{eq:ah2})} \nonumber\\
\sqrt{\frac{M}{M-h} \ln{\Big(\frac{M}{h}\Big)}}
   \quad &{\rm for\ }R\ {\rm in\ Eqn.~(\ref{eq:Rindex})} \nonumber\\
\sqrt[4]{\ln\big(\frac{4M}{\e h}\big)}
   \quad &{\rm for\ }\sqrt{hg}\ {\rm in\ Eqn.~(\ref{eq:hg})} \nonumber\\
\frac{M}{M-h}\ln{\big(\frac{M}{h}\big)}-1
   \quad &{\rm for\ }e^2\ {\rm in\ Eqn.~(\ref{eq:eindex})} \nonumber
\end{align}
\begin{align}
\sqrt{\frac{\frac{M}{M-h}\left(\ln{\frac{M}{h}}\right)-1 }
        {\frac{N}{N-h}\left(\ln{\frac{N}{h}}\right)-1 }}\ h 
   \quad &{\rm for\ }h^\prime\ {\rm in\ Eqn.~(\ref{eq:hprime})} \nonumber\\
\frac{M}{M-h}\ln\frac{M}{h}-1 
   \quad &{\rm for\ }dc_i\ {\rm in\ Eqn.~(\ref{eq:dci1})} \nonumber\\
1-\frac{h}{N-h}\big(\frac{N}{N-h}\ln{(\frac{N}{h})} - 1\big)
   \quad &{\rm for\ }dc_o\ {\rm in\ Eqn.~(\ref{eq:dco1})} \nonumber
\end{align}
are constants in the linear model.

One issue with the $h$-index is that it provides no information for
distinguishing two authors with the same integer value $h$.
Even when they are different, authors may have larger $h$ values
because they have been publishing for a longer time.
Hirsch himself recommended using $h/t$, i.e. $h_0$ in the linear model,
to compare authors with different seniority.

As mentioned above, Burrell used his probabilistic model to examine how
$\Ohead$ varies with time.
If the linear model holds for our approximation, then
\begin{align}
\frac{d\Ohead}{dt}
&\approx 2\Big(\frac{M_0 h_0^2}{M_0 -h_0}\ln{\Big(\frac{M_0}{h_0}\Big)}\Big)t
                           \quad{\rm from\ Eqn.~(\ref{eq:t2})}     \nonumber\\
&= \frac{2}{t}\Big(\frac{Mh^2}{M-h}\ln{\Big(\frac{M}{h}\Big)}\Big) \nonumber\\
&\approx 2\frac{\Ohead}{t} \nonumber
\end{align}
We see that (like using $h/t$ to differentiate two authors with the same $h$)
for two authors with the same $\Ohead$,
the senior author has a larger $t$ and thus
a smaller growth rate $\frac{d\Ohead}{dt}$.

\subsection{$h$-index for a group}
\label{sec:grouph}

The concept of $h$-index has been extended from
an individual author to a group (department~\cite{DeptH2017}, 
journal~\cite{MingersJournalH2008}, etc.).
Here, we apply our equation to derive the $h$-index of a group from the 
individual $h$-indices.

\begin{figure}
\centerline{
\includegraphics[width=8.3cm]{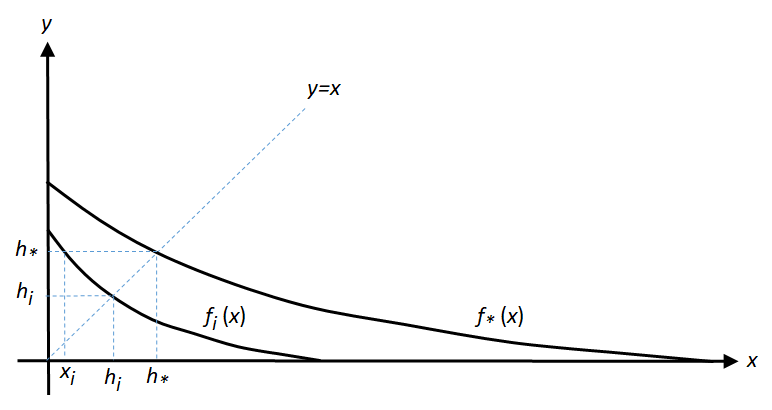}
}
\caption{$h_i\le h_*$, so $x_i\le h_*$.}
\label{fig:hstar0}
\end{figure}

Consider a group of $r$ authors.
Let $M_i$, $N_i$ and $h_i$ be the $M$, $N$ and $h$ values for the $i$-th author
in the group, and $M_*$, $N_*$ and $h_*$ the $M$, $N$ and $h$ values
for the collection of publications from this group.  
Then,
\begin{equation}
M_*=\max\{M_1,\ldots, M_r\}.
\label{eq:Mstar}
\end{equation}
For a first approximation, we assume no two authors in the group share 
a publication, so
\begin{equation}
N_*=N_1+\cdots+N_r .
\label{eq:Nstar}
\end{equation}
Suppose the $i$-th author has $x_i$ publications with at least $h_*$
citations each.
By the definition of the $h$-index,
\begin{equation}
h_*=x_1+\cdots+x_r .
\label{eq:xsum}
\end{equation}
Since $x_i\le h_i$ (see Fig.~\ref{fig:hstar0}),
we can use $\fhead$ to approximate the citation data for each author.
Let $a_i$, $b_i$ and $c_i$ be the $\ahead$, $\bhead$ and $\chead$ values
for the $i$-th author.
By Eqn.~(\ref{eq:abchead}),
\begin{align}
\begin{split}
a_i &= 0 \label{eq:fi} \\
b_i &= \frac{M_i h_i^2}{M_i-h_i} \\
c_i &= \frac{h_i^2}{M_i-h_i}  \\
{\rm and}\quad f_i(x) &= \frac{b_i}{x+c_i}\ ,
\end{split}
\end{align}
where $f_i$ is $\fhead$ for the $i$-th author.
Then
\[
h_* = \frac{b_i}{x_i + c_i} \quad{\rm for\ }i=1,\ldots,r,
\]
so
\[
\sum_{i=1}^r h_* (x_i + c_i) = \sum_{i=1}^r b_i .
\]
By Eqn.~(\ref{eq:xsum}),
\[
h_*^2 + h_*\sum_{i=1}^r c_i - \sum_{i=1}^r b_i = 0.
\]
Thus
\begin{equation}
h_* = \frac{-\sum_{i=1}^r c_i + 
             \sqrt{\big(\sum_{i=1}^r c_i\big)^2 + 4(\sum_{i=1}^r b_i)}}{2}
\label{eq:hstar0}
\end{equation}
Note that $b_i=M_i c_i$ and $M_i>>1$ for most authors.
For $4\sum_{i=1}^r b_i >> (\sum_{i=1}^r c_i )^2$, 
we can use the following approximation:
\begin{align}
h_* &\approx \frac{-\sum_{i=1}^r c_i + \sqrt{4(\sum_{i=1}^r b_i)}}{2}
                                    \nonumber\\
    &\approx\sqrt{\sum_{i=1}^r b_i} \nonumber\\
    &=\sqrt{\sum_{i=1}^r\frac{M_i h_i^2}{M_i - h_i}} . \label{eq:hstar}
\end{align}
To validate this approximation, we selected 9 authors from our dataset.
Table~\ref{tab:hstar} lists, for author $i$, 
the research area, $M_i$, $N_i$ and $h_i$.
For $r=2,\ldots,9$, we grouped the first $r$ authors' publications
to determine the empirical $h_*$ value for the group.

\begin{table*}
\begin{center}
\begin{tabular}{|l|l|r|r|r|}
\hline
 $i$ & research area        & $M_i$ & $N_i$ & $h_i$ \\
\hline
1 & physics                 &  336 &  15 & 13 \\
2 & city planning           &  423 &  90 & 27 \\
3 & public health           & 2108 &  63 & 32 \\
4 & physiology              & 1161 &  34 & 18 \\
5 & computer science        &  262 & 396 & 44 \\
6 & public policy           &  364 & 128 & 31 \\
7 & sociology               &  901 &  64 & 24 \\
8 & psychology              &  272 & 124 & 46 \\
9 & artificial intelligence &  513 &  94 & 19 \\
\hline
\end{tabular}
\qquad
\begin{tabular}{|l|c|c|}
\hline
 & \multicolumn{2}{c|}{$h_*$}                                 \\ \cline{2-3}
$r$ & empirical value & approximation (Eqn.~(\ref{eq:hstar})) \\ \hline
 2  & \ 32            & 30.9 \\
 3  & \ 49            & 44.7 \\
 4  & \ 57            & 48.2 \\
 5  & \ 71            & 68.2 \\
 6  & \ 77            & 75.5 \\
 7  & \ 92            & 79.3 \\
 8  & 103             & 94.0 \\
 9  & 105             & 96.0 \\
\hline
\end{tabular}
\end{center}
\caption{
Comparing approximate value of $h_*$ in Eqn.~(\ref{eq:hstar}) 
to the empirical $h_*$.}
\label{tab:hstar}
\end{table*}

\begin{figure}
\centerline{
\includegraphics[width=8.3cm]{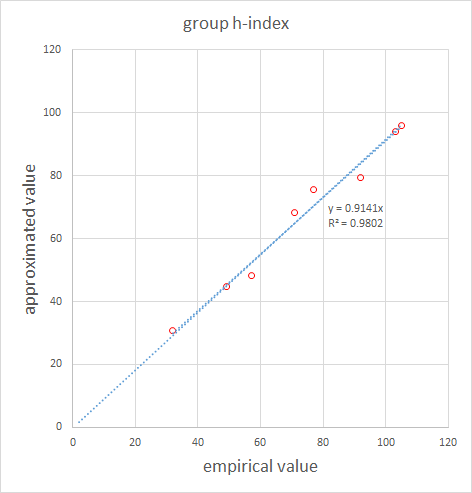}
}
\caption{
Comparing approximate value of $h_*$ in Eqn.~(\ref{eq:hstar}) 
to the empirical $h_*$.}
\label{fig:hstar}
\end{figure}

Note that the approximation (\ref{eq:hstar}) 
takes into account the $M_i$ value for each author in the collection
when estimating the aggregate $h_*$ value.
The omission of $\sum_{i=1}^r c_i$ leads to an under-estimation, 
but Fig.~\ref{fig:hstar} shows that it nonetheless
gives a good approximation for $h_*$
(the regression line has gradient 0.91 and $R^2\approx0.98$).

\section{Conclusion}
\label{sec:conclusion}

In this paper, we proposed a simple equation to approximate the citation count
distribution of an author.
The equation is based on the idea of using 3 geometrical characteristics
($M$, $N$ and $h$) of the count distribution to calibrate 3 parameters 
($a$, $b$ and $c$) for an equation to approximate the distribution.

We demonstrated the equation's usefulness in the analysis of such distributions
by deriving closed-form expressions for various citation indices,
and using them to model the effect of time, and identify individual contribution to a group $h$-index.

\bibliographystyle{abbrv}
\bibliography{cite}{}

\section*{Appendix}

The following histograms describe the sample of 226 authors:

\begin{figure}[h]
\centerline{
\includegraphics[width=8.3cm]{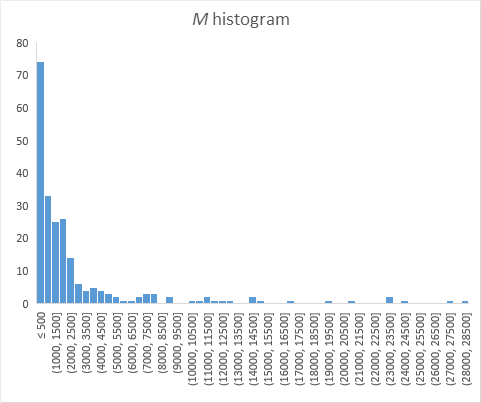}
}
\end{figure}

\begin{figure}[h]
\centerline{
\includegraphics[width=8.3cm]{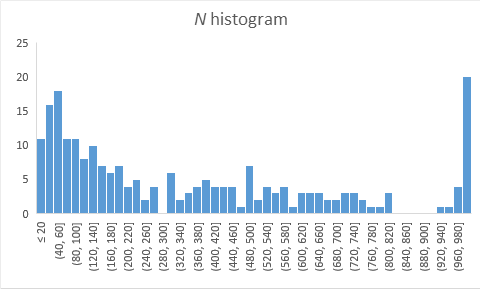}
}
\end{figure}

\begin{figure}[h]
\centerline{
\includegraphics[width=8.3cm]{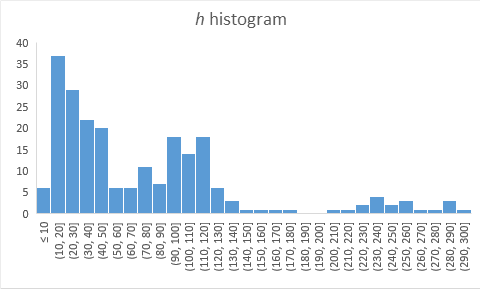}
}
\end{figure}

\end{document}